\documentclass[preprint,aps,prd,superscriptaddress,longbibliography]{revtex4-1}
\pdfoutput=1
\usepackage{bm}
\usepackage{graphicx}
\usepackage{booktabs}
\usepackage{slashed}
\usepackage{hyperref}
\usepackage{amssymb,amsmath,amsfonts}
\usepackage{color}
\usepackage[utf8]{inputenc}
\usepackage[caption=false]{subfig}

\newcommand{\be}{\begin{equation}}
\newcommand{\ee}{\end{equation}}
\newcommand{\bea}{\begin{eqnarray}}
\newcommand{\eea}{\end{eqnarray}}

\renewcommand{\vec}[1]{{\bm #1}}

\begin{document}

\title{How transverse thermal fluctuations disorder a condensate of chiral spirals into a quantum spin liquid}

\author{Robert D. Pisarski}
\affiliation{Department of Physics, Brookhaven National Laboratory, Upton, NY 11973}

\author{Alexei M. Tsvelik}
\affiliation{Condensed Matter Physics and Materials Science Division, Brookhaven National Laboratory, Upton, NY 11973-5000, USA}

\author{Semeon Valgushev}
\affiliation{Department of Physics, Brookhaven National Laboratory, Upton, NY 11973}

\begin{abstract}
  For a scalar theory with a global $O(N)$ symmetry, when $N=2$ a spatially inhomogeneous condensate
  arises when the term in the Lagrangian with two spatial derivatives has a negative coefficient.
  If the condensate for such a chiral spiral includes only one mode, characterized by a momentum $k_0 \hat{z}$,
  then in perturbation theory at nonzero temperature the propagator for the static mode
  has a double pole when $\vec{k}^2 = k_0^2$.  We conjecture that since chiral spirals spontaneously break both
  global and spacetime symmetries, that such double poles are a universal property of their static transverse modes.
  Fluctuations from double poles generate linear infrared divergences in any number of spatial dimensions
  and disorder the condensate of chiral spirals, analogous to a type
  of quantum spin liquid.
  The characteristic feature of this region is that over large spatial distances 
  the two point function is the usual exponential times an oscillatory function.
  We establish this at large $N$ and suggest that it occurs for all $N > 2$.
  Implications for fermion models and the phase diagram of QCD at nonzero density are discussed.
\end{abstract}

\maketitle

Scalar field theories with a global $O(N)$ symmetry are a useful paradigm for the study of phase transitions
\cite{chaikin:2010,Fradkin:1991nr}.  In
vacuum there are two possible phases: symmetric, when the mass squared is positive, and broken, when it is negative.
The latter spontaneously breaks the $O(N)$ symmetry and generates Goldstone bosons.

At nonzero temperature or net density Lorentz symmetry is lost, and a different phase can arise.
By causality, terms with two time derivatives must always have a positive coefficient.  However, 
for an effective Lagrangian in a medium the coefficient of the term quadratic in the spatial derivatives, $Z$,
can be negative.  Of course, stability must be ensured by adding terms with four or more spatial derivatives with
positive coefficients.
When $Z < 0$, spatially inhomogeneous condensates arise naturally, as a balance between the negative term
with two spatial derivatives against those with higher powers \cite{chaikin:2010,Fradkin:1991nr,Pisarski:2018bct}.
These appear in many systems, from smectic liquid crystals \cite{chaikin:2010,Fradkin:1991nr},
to inhomogeneous polymers \cite{Fredrickson:2006}, to complex fluids \cite{Cates12}, to chiral spirals in pions and kaons
\cite{Pisarski:2018bct,Overhauser:1960,Migdal:1971,Migdal:1973zm,Migdal:1978az,Kleinert:1981zv,Kolehmainen:1982jn,Baym:1982ca,Bunatian:1983dq,Takatsuka:1987tn,Migdal:1990vm,Kaplan:1986yq,Brown:1993yv,Shei:1976mn,Thies:2006ti,Basar:2008im,Basar:2008ki,Basar:2009fg,Kojo:2009ha,Kojo:2010fe,Kojo:2011cn,Kojo:2014fxa,Nickel:2009ke,Nickel:2009wj,Buballa:2014tba,Carignano:2014jla,Hidaka:2015xza,Lee:2015bva,Buballa:2015awa,Braun:2015fva,Carignano:2015kda,Heinz:2015lua,Carignano:2016jnw,Azaria:2016mqb,Adhikari:2016jzc,Adhikari:2016vuu,Andersen:2017lre,Adhikari:2017ydi,James:2017cpc,Carignano:2016lxe,Khunjua:2017khh,Khunjua:2017mkc,Andersen:2018osr,Carignano:2018hvn,Buballa:2018hux,Khunjua:2018sro,Khunjua:2018jmn,Carignano:2019ivp,Khunjua:2019lbv,Khunjua:2019ini,Thies:2019ejd,Thies:2020gfy,Pannullo:2019bfn,Pannullo:2019prx,Lenz:2020bxk,Narayanan:2020uqt}.

It is well known that spatially inhomogeneous condensates exhibit a variety of infrared divergences.  Due to the anisotropic
propagator of a longitudinal phonon mode, they are disordered logarithmically over large distances
\cite{chaikin:2010,Fradkin:1991nr,Pisarski:2018bct,Kolehmainen:1982jn,Baym:1982ca,Bunatian:1983dq,Takatsuka:1987tn,Migdal:1990vm}.
Fluctuations also ensure that the transition between the symmetric and spatially inhomogeneous
phase is of first, and never of second, order
\cite{Pisarski:2018bct,Brazovski:1975,Ling:1981,Dyugaev:1982gf,Hohenberg:1995,Karasawa:2016bok}.
Lastly, at or below four spatial dimensions infrared divergences prevent the appearance of
a Lifshitz point, where the mass squared and $Z$ vanish simultaneously
\cite{Fradkin:1991nr,Erzan:1977,Sak:1978,Grest:1978,Fredrickson:2006,Bonanno:2014yia,Zappala:2017vjf,Zappala:2018khg,Pisarski:2018bct}.

In the broken phase, the condensate is constant in both the internal and coordinate space,
and generates $N-1$ transverse massless modes.
When $Z < 0$ and $N=2$ the condensate consists of the two fields which rotate into one another as one moves
along a fixed direction in space.  
In this paper we consider how the $N-2$ transverse modes affect spatially inhomogeneous condensates when $N > 2$.

In Sec. (\ref{sec:pert}) we consider a perturbative analysis of the transverse modes.  The effective Lagrangian we
consider is given in Sec. (\ref{sec:eff_lag}).  In Sec. (\ref{sec:single}) we consider the ansatz for a chiral spiral
with a single mode, characterized by a momentum along a given direction, $k_0 \hat{z}$.
We show that for a specific choice of Lagrangian, the static mode has a double pole at nonzero
momentum, for $\vec{k}^2 = k_0^2$.  We generalize this to a wide class of effective Lagrangians in Sec. (\ref{sec:general}).
While our analysis is limited to the ansatz of a single mode for the condensate, we suggest that it is generic.

Since it occurs at nonzero momentum,
such a double pole produces a severe, linear infrared divergence even in the simplest
tadpole diagram.  This suggests that the theory is in a novel disordered state.  This can be established at large $N$,
where the theory is soluble by standard techniques
\cite{Tsvelik:1996zj,Moshe:2003xn}.  We give a general discussion of the expansion at large $N$ in Sec.
(\ref{sec:gen_largeN}), and an explicit solution in Sec. (\ref{sec:soln_largeN}).  
This shows that there are only two phases, broken and symmetric.
Even so, there are two parts to the symmetric phase: one which is ordinary, and 
one which we refer to as a type of ``quantum spin liquid''.  While there is
no phase transition between the ordinary symmetric phase and the quantum spin liquid, it is easy
to distinguish between the two, as the two point function in the ordinary symmetric phase
is just the usual exponential(s), while that in the quantum spin liquid is an exponential times 
an oscillatory function.
The quantum spin liquid includes the entire region where mean field theory indicates
a condensate of chiral spirals, plus a larger part of the parameter space.  

In Sec. (\ref{sec:phase_diagram}) we conjecture that static, transverse modes produce a quantum spin liquid for all $N > 2$.
This is contrast to Kleinert \cite{Kleinert:1981zv}, who added a term to the effective
Lagrangian which is nonanalytic in the momenta.
This generates a double pole and so disorder for all $N \geq 1$
\footnote{In a spatially inhomogeneous condensate along the $z$ direction,
  the inverse propagator for the longitudinal mode is anistropic, 
  $\sim (k_z - k_0)^2 + \ldots$, Eq. (\ref{long_prop})
  \cite{chaikin:2010,Fradkin:1991nr,Pisarski:2018bct,Kolehmainen:1982jn,Baym:1982ca,Bunatian:1983dq,Takatsuka:1987tn,Migdal:1990vm,Brazovski:1975,Ling:1981,Dyugaev:1982gf,Hohenberg:1995,Karasawa:2016bok,Lee:2015bva,Hidaka:2015xza,Nitta:2017mgk,Nitta:2017yuf,Gudnason:2018bqb}.  A term linear in $k_z$, $\sim - 2 k_0 k_z$,
  is allowed because the condensate spontaneously breaks the rotational symmetry.  
  Instead, Eq. (5) of Ref. \cite{Kleinert:1981zv} introduces a term $\sim - k_0 \sqrt{\vec{k}^{2}}$.  This does not
  involve the direction of the condensate and so unlike ordinary effective Lagrangians, is not analytic in $\vec{k}^{2}$.
  This term generates a double pole at nonzero momentum for
  both the longitudinal and transverse modes and so disorders for all $N$; we only find disorder for $N > 2$.
  Similar to our analysis, Ref. \cite{Kleinert:1981zv} finds that the double pole is absent at zero temperature,
  as with our Eq. (\ref{zero_temperature}); the model is solved at large $N$, similar to our Sec. (\ref{sec:soln_largeN}),
  although since the model differs, so do the details. More generally, we argue
  for a double pole to arise from an effective Lagrangian which has a well defined derivative expansion,
  then it {\it must} involve higher spatial derivatives.},
in contradiction to
numerous systems in condensed matter which exhibit quasi-long-range order for
$N=1$ and $2$ \cite{chaikin:2010,Fradkin:1991nr,Pisarski:2018bct,Fredrickson:2006,Cates12}.
Lee, Nakano, Tatsumi, Tsue, and Friman \cite{Lee:2015bva} analyzed an $O(4)$ model, but did not find
double poles for the static transverse modes
\footnote{Ref. \cite{Lee:2015bva} consider fluctuations which multiply 
  the background field,
  $\vec{\phi} = (\vec{\phi}_0 + \sigma_q) \exp( i t^a \phi^a_q(x))$,
  where $t^q$ are generators of $O(4)$.  The first term on the right
  hand side of their Eq. (15) does exhibit a double pole in terms of their $\vec{\beta}_U
  = \vec{\beta}_T \, cos(k_0 z)$, but the spectrum of their $\vec{\beta}_T$ does not appear to exhibit a double pole,
  Eqs. (16)-(20) and Appendix B of Ref. \cite{Lee:2015bva}.
  We expand in linear fluctuations about the background field,
  $\vec{\phi} = (\vec{\phi}_0 + \vec{\sigma}_q, \vec{\chi})$, our Eq. (\ref{decompose_two}),
  where $\vec{\chi}$ is related to their $\vec{\beta}_U$
  and $\vec{\beta}_T$.  There is a double pole at nonzero momentum in the static inverse
  propagator for $\vec{\chi}$, Eqs. (\ref{inverse_trans}), (\ref{inverse_trans_mass}),
  (\ref{inverse_trans_prop_1stderiv}), and (\ref{inverse_chi_prop_solve}). }.
The nature of Goldstone bosons for spatially inhomogeneous condensates has also been studied by
Hidaka, Kamikao, Kanazawa, and Noumi \cite{Hidaka:2015xza} in an NJL model
\footnote{
Ref. \cite{Hidaka:2015xza} considers a kink crystal in a
Nambu Jona-Lasino (NJL) model in 3+1 dimensions, and find anisotropic Goldstone modes
about zero momentum. As discussed in Sec. \ref{sec:NJL}, the NJL model only breaks
a global $Z(2)$ symmetry.  As we discuss in Sec. \ref{sec:single}, we expect that the spontaneous
breaking of an O(4) symmetry prefers a chiral spiral over a kink crystal.}
and for $N=2$ by Gudnason, Nitta, Sasaki, and Yokokura \cite{Nitta:2017mgk,Nitta:2017yuf,Gudnason:2018bqb}.

In Sec. (\ref{sec:phase_diagram})
we also discuss the modifications of the phase diagram from mean field theory \cite{Pisarski:2018bct}, and
what happens when the $O(N)$ symmetry is only approximate.  We show that
the linear infrared divergences of the transverse modes only appear 
at nonzero temperature, while at zero temperature the infrared divergences are only logarithmic \cite{Pisarski:2018bct}.  

Admittedly, the analogy to a quantum spin liquid in condensed matter systems
\cite{Fradkin:1991nr,Ioffe:1988a,Zhou:2016cnl,Savary:2016ksw,Kharkov:2017cvm,Kharkov:2018lri,Kharkov:2019fvv,OBrien:2020bho}
is imprecise.
Usually, the disorder in a quantum spin liquid arises due to frustration, such as for an antiferromagnet
on a triangular lattice in two dimensions.  As such, it automatically
persists at zero temperature.  In contrast, our model exhibits
quasi-long-range order at zero temperature.

Sec. (\ref{sec:NJL}) proposes a fermion model in $2+1$ dimensions
which could be used to test our predictions at nonzero chemical potential.  Sec. (\ref{sec:QCD}) 
briefly considers implications for the phase diagram of Quantum ChromoDynamics (QCD)
\cite{Asakawa:1989bq,Stephanov:1998dy,Stephanov:1999zu,Son:2004iv,Stephanov:2008qz,Parotto:2018pwx,Schaefer:2006ds,Rennecke:2016tkm,Fu:2019hdw,Bzdak:2019pkr}.

\section{Perturbative analysis}
\label{sec:pert}

\subsection{Effective Lagrangian}
\label{sec:eff_lag}

Consider the usual form for the effective Lagrangian,
\begin{equation}
  {\cal L}_0 = \frac{1}{2} \left( \partial_0 \vec{\phi}\right)^2 
  + \frac{Z}{2} \left( \partial_i \vec{\phi}\right)^2 - h \vec{\phi} \cdot \vec{\phi}_b + \frac{1}{2} m^2 \vec{\phi}^{\, 2}
  + \frac{1}{4} \lambda (\vec{\phi}^{\, 2})^2 
  + \frac{1}{6} \kappa (\vec{\phi}^{\, 2})^3 \; .
  \label{lag0}
\end{equation}
Here $\vec{\phi}$ is an $N$-component vector, and so the theory has a global symmetry of $O(N)$.
The background field, $h$, violates the $O(N)$ symmetry; we usually assume it vanishes, and comment
briefly what happens when it is nonzero but small.  
We assume that
the theory is applicable in a medium, and so Euclidean invariance need not apply.  This allows for the coefficient of
the term with two spatial derivatives, $Z$, to differ from that with two time derivatives.

We not only allow $Z \neq 1$ in a medium, but let $Z$ to be negative.  In that case, it is necessary to add
terms with more spatial derivatives in order to stabilize the theory,
\begin{equation}
{\cal L}_{\rm HD} = \frac{1}{2 M^2} \left( \partial_i^2 \vec{\phi}\right)^2
+ \frac{1}{2 M_1} \vec{\phi}^{\, 2}  \left( \partial_i \vec{\phi}\right)^2 + \frac{1}{2 M_2} (\partial_i \vec{\phi}^{\, 2})^2 \; .
\label{laghd}
\end{equation}

We consider a theory in $d+1$ dimensions at a nonzero temperature $T$, so the energy
$E_n = 2 \pi T n$, $n = 0, \pm 1, \pm 2 \ldots$.  In this case, the most infrared divergent mode is
the static mode with zero energy, $E_n = 0$, and the effective theory is that for the static mode in
$d$ dimensions.  We comment later in Sec. (\ref{sec:phase_diagram}) how the infrared divergences are less
severe at zero temperarture.  This is obvious, as at zero temperature the integral over a continuous energy,
$E$, smooths out the infrared divergences from the spatial momenta of the transverse mode.

In three spatial dimensions, $d=3$, $\vec{\phi}$ has dimensions of ${\rm mass}^{1/2}$.  The terms in Eq. (\ref{laghd})
are all non-renormalizable couplings, where $M$, $M_1$, and $M_2$ are characteristic
of some large mass scale generated by the medium.  At zero temperature, such as in a medium at nonzero density,
$\vec{\phi}$ has dimensions of mass, and the analogous couplings are $1/M^2$, $1/M_1^2$, and $1/M_2^2$, respectively.

Most of our analysis is not sensitive to the number of spatial dimensions.  However, we note that in
one spatial dimensions, $d=1$, $\vec{\phi}$ is dimensionless.  In that case, it is also natural to use a 
nonlinear Lagrangian, where $\vec{\phi}^2 = 1$.  The nonlinear Lagrangian is
\begin{equation}
  {\cal L}_{2 \; {\rm dim}} = \frac{1}{2} (\partial_0 \vec{\phi})^2 + \frac{Z}{2} (\partial_i \vec{\phi})^2
  + \frac{1}{2 M^2} (\partial^2 \vec{\phi})^2 + \frac{1}{2 M_2^2} ( (\partial_i \vec{\phi})^2 )^2 \; .
\end{equation}
Because of the constraint there is no term $\sim 1/M_1$, but there still are two different terms with four
spatial derivatives, $\sim 1/M^2$ and $\sim 1/M_2^2$.

\subsection{Single mode ansatz}
\label{sec:single}

Henceforth we concentrate on the static mode at nonzero temperature, in zero background field, $h = 0$.
The condensate we consider arises when two conditions are met: first, a negative mass squared, $m^2 < 0$.
In mean field theory, a negative mass term is balanced by a positive term for the quartic coupling (or of higher
order, if that is negative).  This balancing 
generates a nonzero expectation value for the field $\langle \vec{\phi} \rangle = \phi_0 \, \hat{\vec{\phi}}$,
for some fixed direction $\hat{\vec{\phi}}$.  For $N > 1$, the choice of $\hat{\vec{\phi}}$
spontaneously breaks the $O(N)$ symmetry.
If $N=1$, the only choice is the sign of the condensate, $\langle \phi \rangle = \pm |\phi_0|$, which spontaneously
breaks a global $Z(2)$ symmetry.

If the coefficient of the term with two spatial derivatives is negative, $Z < 0$, a similar balancing occurs.
The theory generates a condensate where $\langle (\partial_i \vec{\phi} )^2 \rangle$ is nonzero, with terms with
more derivatives, Eq. (\ref{laghd}),
stabilizing the theory.  If $\partial_i \vec{\phi} \neq 0$, this condensate is spatially inhomogeneous.  We always
assume that the inhomogeneity is always only in one direction.  Condensates in several directions are possible,
but typically have higher energy.

The detailed form of the spatially inhomogeneous condensate depends crucially upon $N$.  The equation of
motion is, from Eqs. (\ref{lag0}) and (\ref{laghd}),
\begin{eqnarray}
  \frac{1}{M^2} (-\partial_i^2)^2 \vec{\phi} &+& \frac{1}{M_1} \left( (\partial_i \vec{\phi})^2 \vec{\phi} -
    \partial_i (\vec{\phi}^{\, 2} \partial_i \vec{\phi} ) \right)
                   + \frac{2}{M_2} \partial_i ( (\partial_i \vec{\phi}^{\, 2}) \vec{\phi} ) 
   + Z (-\partial_i^2) \vec{\phi}                \nonumber \\
     &+& \left( m^2 + \lambda \vec{\phi}^{\, 2} + \kappa (\vec{\phi}^{\, 2})^2 \right) \vec{\phi} = 0 \; .
 \label{eom}
\end{eqnarray}

When $Z$ is positive, the theory generates a constant condensate, $\phi_0 = \pm \sqrt{- m^2/\lambda}$
when $\kappa = 0$.  When $N=1$, the only way 
to develop $\langle (\partial_i \vec{\phi} )^2 \rangle \neq 0$
is for the field to oscillate between the vacuum values, $\pm |\phi_0|$,
with some periodicity.  This ``kink'' crystal is a solution
of the nonlinear differential equation in Eq. (\ref{eom}).  For the 
Gross-Neveu model in $1+1$ dimensions the precise form of the kink crystal can be computed
analytically at large $N$, and is not trivial \cite{Shei:1976mn,Thies:2006ti,Thies:2019ejd,Thies:2020gfy}.
By using non-Abelian bosonization and the truncated conformal spectrum approach,
a generalized Gross-Neveu model, with two flavors and three colors, is also exactly soluble in
$1+1$ dimensions \cite{Azaria:2016mqb,James:2017cpc}.  

When $N \geq 2$, however, there is a {\it much} simpler ansatz which generates a condensate for both for $\phi$ {\it and}
its spatial derivatives.  The magnitude of $\phi$ is kept constant, with the spatial derivative generated by
a rotation in the internal space:
\begin{equation}
  \vec{\phi}_0(\vec{x}) = \phi_0 (\cos(k_0 z) , \sin(k_0 z), \vec{0} ) \; .
  \label{cs_ansatz}
\end{equation}
This is periodic is some fixed direction, $z$, with period $2 \pi/k_0$.  This choice of the direction $z$
spontaneously breaks the rotation symmetry.

More general solutions are certainly possible.  That of Eq. (\ref{cs_ansatz}) involves only a single
mode.  The most general solution with this periodicity involves an infinite number of modes:
\begin{equation}
   \vec{\phi}_0(\vec{x}) = (\sum_{n = 1}^\infty c_n \cos(n \, k_0 z) , \sum_{n = 1}^\infty s_n \sin(n \, k_0 z), \vec{0} ) \; .
  \label{cs_ansatz_multi}
\end{equation}
The great advantage of the single mode ansatz of Eq. (\ref{cs_ansatz}) is that $\vec{\phi}^{\, 2}$ is a constant,
which makes solving the equations of motion trivial.  In constrast, for
the multi-mode solution of Eq. (\ref{cs_ansatz_multi}) $\vec{\phi}^2$ is not constant, and
there are an infinite number of parameters which need to
be determined, the coefficients $c_n$ and $s_n$.  We only analyze the single mode solution of Eq. (\ref{cs_ansatz}),
assuming that it is the vacuum of our model.  
We comment that for the chiral Gross-Neveu model in $1+1$ dimensions, at large N the chiral spiral is computable
analytically, and is a multi-mode solution \cite{Basar:2008im,Basar:2008ki,Basar:2009fg}.  This theory is rather
more involved than the elementary scalar field theory which we analyze.  In any case, we speculate that even if
the single mode solution is not the vacuum, that the most interesting properties remain valid for the multi-mode solution.

With the single mode ansatz of Eq. (\ref{cs_ansatz}), the Lagrangian equals
\begin{equation}
  {\cal L} = \frac{1}{2 M^2} \; k_0^4 \; \phi_0^2 + \frac{1}{2 M_1} \; k_0^2 \; \phi_0^4
  + \frac{Z }{2} \; k_0^2 \; \phi_0^2
  + \frac{m^2}{2} \; \phi_0^2 + \frac{\lambda}{4} \; \phi_0^4 + \frac{\kappa}{6} \; \phi_0^6 \; .
\end{equation}
We first vary with respect to $k_0$,
\begin{equation}
  \frac{\partial }{\partial k_0} {\cal L} = 0 \; .
  \label{vary_k0}
\end{equation}
This is equivalent to minimizing the energy per period of the condensate \cite{Hidaka:2015xza}.
Because our ansatz has constant $\vec{\phi}^2$, it is independent of $M_2$.  The solution is
\begin{equation}
  k_0^2 = \left( - Z - \frac{\phi_0^2}{M_1} \right) \; \frac{M^2}{2} \; .
  \label{soln_k0}
\end{equation}
$k_0^2$ has to be positive, so this equation can be satisfied if
$Z$ is sufficiently large and negative.  Under this condition, substituting $k_0$ back into the
Lagrangian gives
\begin{equation}
  {\cal L}(k_0, \phi_0) = \frac{1}{2} \left( m^2 - \frac{Z^2}{4} \, M^2 \right) \phi_0^2
  + \frac{1}{4} \left( \lambda -  Z \, \frac{M^2}{M_1} \right) \phi_0^4 
  + \frac{1}{6} \left( \kappa - \frac{3}{4} \, \frac{M^2}{M_1^2} \right) \phi_0^6 \; .
  \label{eom_phi0}
\end{equation}
The equation of motion for the vacuum expectation value is then determined by the solution of
\begin{equation}
  \frac{\partial {\cal L}}{\partial \phi_0} = \left( m^2 - \frac{Z^2}{4} \, M^2 \right) \phi_0
  + \left( \lambda -  Z \, \frac{M^2}{M_1} \right) \phi_0^3
  + \left( \kappa - \frac{3}{4} \, \frac{M_2^2}{M_1^2} \right) \phi_0^5 = 0
  \; .
\end{equation}
When a spatially inhomogeneous condensate develops, $Z < 0$ and $k_0 \neq 0$, this affects the couplings of the
scalar potential, including both the mass squared, and the quartic and hexatic couplings.

The propagator for the longitudinal modes is involved, and involves a phonon mode associated with the spontaneous
breaking of translational symmetry, along $z$, by the condensate
\cite{chaikin:2010,Fradkin:1991nr,Pisarski:2018bct,Lee:2015bva,Hidaka:2015xza,Nitta:2017mgk,Nitta:2017yuf,Gudnason:2018bqb}.

The propagator for the transverse modes is easy to compute, though
\begin{equation}
  \vec{\phi} = (\vec{\sigma},\vec{\chi}) \; .
\end{equation}
Then the inverse propagator for a static $\vec{\chi}$ field is
\begin{equation}
  \Delta_\chi^{-1}(E = 0, \vec{k})
  = \frac{1}{ M^2} \, (\vec{k}^{\, 2})^2 + \left( Z + \frac{\phi_0^2}{M_1} \right)\vec{k}^{\, 2}
  + \frac{1}{M_1} \, k_0^2 \, \phi_0^2 + m^2 + \lambda \, \phi_0^2 + \kappa \phi_0^4 \; .
\end{equation}
Using the expression for $k_0$, Eq. (\ref{soln_k0}), 
\begin{equation}
  \Delta_\chi^{-1}(0,\vec{k}) = \frac{1}{M^2} \left( \vec{k}^{\, 2} - k_0^2 \right)^2 + {\cal M}^2 \; .
  \label{inverse_trans}
\end{equation}
After some algebra, we find that
\begin{equation}
  {\cal M}^2 = \frac{1}{\phi_0} \; \frac{\partial {\cal L}}{\partial \phi_0} = 0 \; .
  \label{inverse_trans_mass}
\end{equation}
By direct computation it is not obvious that the mass squared, about $k_0$, is proportional to the equation
of motion, and so vanishes.  We show this for a very general model in the next section,
Eq. (\ref{inverse_trans_prop_k0}).

That the mass squared vanishes at zero momentum for the transverse modes is simply an 
expression of Goldstone's theorem.  That is does so for a chiral spiral with a single mode,
about the characteristic momentum $k_0$ of the condensate, is not obvious.

Further, while the transverse modes in Eq. (\ref{inverse_trans}) are massless at $k_0$,
they do so through a {\it double} pole, which generates severe infrared divergences.
Any tadpole diagram involving the transverse field is proportional to
\begin{equation}
  \int d^d \vec{k} \; \frac{1}{(\vec{k}^2 - k_0^2)^2/M^2 + {\cal M}^2}
  \sim \frac{M^2}{k_0^2} \; \int_{k \sim k_0} \frac{dk}{(k - k_0)^2} \; .
  \label{tadpole_div}
\end{equation}
In any number of dimensions, including $d=1$,
this is a {\it linear} infrared divergence about $k_0$.  Consequently, the tadpole diagram blows up, and the ansatz is
certainly destablized once quantum fluctuations are included.

We find that this linear infrared divergence from the double pole in the transverse fluctuations again when we analyze the
model at large $N$ in Sec. (\ref{sec:gen_largeN}).  The advantage of analyzing the model at large $N$ is that then
we can be sure that this infrared divergence is not cut off by other diagrams, or other effects which we might miss
in a perturbative analysis.

\subsection{General Lagrangian for the single mode ansatz}
\label{sec:general}
This result can be generalized to the following Lagrangian:
\begin{equation}
  {\cal L} = \frac{1}{2} \left( \partial_0 \vec{\phi}\right)^2
  + \sum_{n=1}^{\infty}  \frac{Z_n}{2} \vec{\phi} (-\partial^2)^n \vec{\phi}
  + \frac{v_n}{2} (\vec\phi^2)^n \; .
\label{gen_lag}
\end{equation}

This assumes {\it arbitrary} powers of spatial derivatives and of $\vec\phi^2$ in the potential.                                    
We do assume that terms with spatial derivatives only include two powers of $\vec\phi$, and not                                     
higher powers of $\phi$, such as                                                                                                    
$\vec{\phi}^{\, 2}  ( \partial_i \vec{\phi})^2$ and $ (\partial_i \vec{\phi}^{\, 2})^2$                                             
in Eq. (\ref{laghd}).  We suspect that such terms could be included, but as
the number of spatial                                       
derivatives increases, though, so does the number of such terms, and so we simply ignore
these to emphasize the physics.

What is essential is that our ansatz involves only a single mode in momentum space,
Eq. (\ref{cs_ansatz}), so that $\vec\phi^2$ is of constant magnitude.  With this ansatz, for static
fields the Lagrangian becomes
\begin{equation}
  {\cal L}(k_0,\phi_0) = \frac{1}{2} \sum_{n=1}^\infty Z_n \, k_0^{2n} \, \phi_0^2 + v_n \phi_0^{2 n} \; .
\end{equation}

Varying with respect to $\phi_0$ gives 
\begin{equation}
  \frac{\partial {\cal L}}{\partial \phi_0} = \sum_{n=1}^\infty \left(
    Z_n \, k_0^{2 n} \; + n \ v_n \phi_0^{2n - 2} \right) \phi_0 = 0 \; .
  \label{vary_phi0}
\end{equation}

The solution is either $\phi_0 = 0$ or a nonzero value of $\phi_0$.  The solutions to Eqs. (\ref{vary_k0})
and (\ref{vary_phi0}) are, in general, involved.

However, if the {\it only} thing we wish to do is to calculate the transverse propagator, we do not require the
explicit form of the solution.  This depends crucially upon the point that with a single mode ansatz, $\vec\phi^2$
is constant.  The inverse propagator for the transverse modes is just
\begin{equation}
\Delta^{-1}(\vec{k}) = \sum_{n=1}^\infty Z_n (\vec{k}^2)^n + n \, v_n \, \phi_0^{2n - 2} \; .                                
\label{inverse_trans_prop}
\end{equation}
Consider the value of this propagator at a point $\vec{k} = \hat{k} k_0$, where $\hat{k}^2 = 1$ is a unit
vector.  That is, the momentum must have magnitude $k_0$, but need not lie along the $z$ direction of the
condensate.  Then                                                                                                                   
\begin{equation}
  \Delta^{-1}(k_0 \, \hat{k} ) = \sum_{n=1}^\infty Z_n k_0^{2n} + n \, v_n \, \phi_0^{2n - 2} \; .
  \label{inverse_trans_prop_k0}
\end{equation}
By the equation of motion for $\vec\phi$, Eq. (\ref{vary_phi0}), this {\it vanishes}.  Thus the
transverse modes have a zero at $\vec{k} = k_0 \hat{k}$.       
Next, expand the inverse transverse propagator about this point:
\begin{equation}
  \left.  \frac{\partial}{\partial k} \Delta^{-1}(\vec{k})\right|_{\vec{k} = \hat{k} k_0}
  = 2 \sum_{n=1}^\infty n \, Z_n \, (k_0)^{2n - 1} \hat{k} \; .
  \label{inverse_trans_prop_1stderiv}
\end{equation}
But by the stationary point condition with respect to $k_0$, Eq. (\ref{vary_k0}), this also vanishes.
                                                                                                                                    
The second derivative of the inverse transverse propagator about $k_0 \hat{k}$ is
\begin{equation}
  \left.  \frac{\partial^2}{\partial k^2} \Delta^{-1}(\vec{k})\right|_{\vec{k} = k_0 \hat{k}}
  = 2 \sum_{n=1}^\infty n (2n-1) \, Z_n \, (k_0)^{2n - 2}  \; .
  \label{inverse_trans_prop_2nd_deriv}
\end{equation}
There is no reason for this quantity to vanish.  Indeed, neither can we be certain that it is positive.
However, because of the coefficient $\sim n(2n-1)$, the terms with the highest $n$ dominate, and it is
reasonable to assume so.  The positivty of Eq. (\ref{inverse_trans_prop_2nd_deriv}) is, in any case,
a necessary condition for stability of the theory.

Since the first derivative of the propagator vanishes at $\hat{k} k_0$, and the second does not,
this establishes the existence of a double zero, at nonzero momentum, for an extremely general form
of the effective Lagrangian.

It is possible that the infrared divergence of the single mode solution is eliminated by going to a multi-mode solution,
Eq. (\ref{cs_ansatz_multi}).  As mentioned previously, we expect that the vacuum with the lowest energy is that
with a single mode.  Even if this is not true, we suggest that the appearance of
double poles in the static, transverse propagator is generic, following from the spontaneous breaking of
both the internal and spacetime symmetries.  
Presumably the double pole is about the smallest periodic momentum, $k_0$.  

\section{General analysis at large $N$}
\label{sec:gen_largeN}

\subsection{Positive $Z$}
\label{sec:gen_posZ}

We treat both symmetric and broken phases simultaneously.  For a large $N$ expansion \cite{Tsvelik:1996zj,Moshe:2003xn}, we take
\begin{equation}
  \vec{\phi} = (\sigma,\vec{\chi}) \; ,
\end{equation}
When the symmetry breaks, we assume that $\langle \sigma \rangle \neq 0$.
We integrate out the $N-1$ component field $\vec{\chi}$ by introducing a constraint field
\begin{equation}
  {\cal L}_{{\rm cons}} = \frac{i \epsilon}{2} (\omega - \sigma^2 - \vec{\chi}^{\, 2}) \; .
  \label{constraint_one}
\end{equation}
Integrating out the $\vec{\chi}$ field, the effective action is
\begin{eqnarray}
  {\cal S} = \int d^3 x \left(
    \frac{1}{2 M^2} (\partial^2 \sigma)^2 \right.
    &+&  \left. \frac{Z}{2} (\partial_i \sigma)^2 + {\cal V}(\omega) 
    + \frac{i \epsilon}{2} (\omega - \sigma^2) \right) \nonumber \\
  &+& \frac{(N-1)}{2} \, {\rm tr} \log \left(  \frac{1}{M^2} (-\partial^2 )^2 
    + Z (- \partial^2) + i \epsilon \right) \; .\\
\end{eqnarray}
For simplicity we assume that $M_1 = M_2 = \infty$ and $\kappa = 0$, but it is direct to include them.

Assume that the stationary point for $\sigma$ is constant.  Denoting the stationary points by
$\sigma_0$, $\epsilon_0$, and $\omega_0$, the equation of motions give:
\begin{equation}
  \epsilon_0 \, \sigma_0 = 0 \; ;
  \label{sigma_eom1}
\end{equation}
for $\sigma$,
\begin{equation}
  \frac{i}{2} \epsilon_0 + \left. \frac{\partial}{\partial \omega} {\cal V}(\omega)\right|_{\omega = \omega_0} = 0 \; ,
  \label{omega_eom}
\end{equation}
for $\omega$, and
\begin{equation}
  \frac{1}{2} (\omega_0 - \sigma_0^2)
  + \frac{(N-1)}{2} \, {\rm tr} \frac{1}{(-\partial^2 )^2/M^2 + Z (- \partial^2) + i \epsilon_0 } = 0 \; ,
  \label{eps_eom}
\end{equation}
for $\epsilon$.

The equation of motion in Eq. (\ref{sigma_eom1}) is especially useful.  In the broken phase $\sigma_0 \neq 0$,
and so $\epsilon_0$ vanishes.  From Eq. (\ref{eps_eom}), this ensures that the transverse modes are Goldstone
bosons.  Eq. (\ref{omega_eom}) determines $\omega_0$, which necessarily has a nonzero value.

Conversely, in the symmetric phase $\sigma_0$ vanishes, and so $\epsilon_0$ is nonzero, determined by
Eq. (\ref{omega_eom}).  From Eq. (\ref{eps_eom}), the transverse modes are massive.  With some computation,
it can be shown that they are degenerate with the longitudinal mode.

\subsection{$ Z < 0$}
\label{sec:gen_negZ}

We now decompose $\vec{\phi}$ into $\vec{\sigma}$ and a $N-2$ component vector, $\vec{\chi}$,
\begin{equation}
  \vec{\phi} = (\vec{\sigma},\vec{\chi}) \; .
  \label{decompose_two}
\end{equation}
We introduce two constraint fields:
\begin{equation}
  {\cal L}_{\rm cons} = \frac{i \, \epsilon}{2} \left( \omega - \vec{\chi}^2 \right)
  + \frac{i \, \widetilde{\epsilon}}{2} \left( \widetilde{\omega} - (\partial_i \vec{\chi})^2 \right)
  \label{constraint_two}
\end{equation}
Here we only introduce constraint fields for the transverse fluctuations, for both their magnitude and
the square of their derivative.  This complicates the form of the effective Lagrangian, but ensures
that we isolate the dynamics of the these modes.

We set $M_1 = M_2 = \kappa = 0$, but it is immediate to generalize our results.
After integrating out $\vec{\chi}$, the effective Lagrangian is
\begin{eqnarray}
  {\cal S}_{\rm eff} = \int d^3 x \left(
    \frac{1}{2 M^2} (\partial^2 \vec{\sigma})^2 \right. &+& \frac{Z}{2} (\partial_i \vec{\sigma})^2 
   + \frac{m^2}{2} \vec{\sigma}^2 + \frac{\lambda}{2} \omega \vec{\sigma}^2
                           + \frac{\lambda}{4} (\vec{\sigma}^2)^2 \nonumber \\
  &-&\left. \frac{i}{2} \left( \epsilon \, \omega + \widetilde{\epsilon} \; \widetilde{\omega} \right)
  + \frac{Z}{2} \, \widetilde{\omega} + \frac{m^2}{2} \, \omega + \frac{\lambda}{4} \, \omega^2 \right) \nonumber \\
  &+& \frac{(N-2)}{2} \, {\rm tr} \log \left(  \frac{1}{M^2} (-\partial^2 )^2 
    - i \partial_i \left( \widetilde{\epsilon} \, \partial_i \right) + i \epsilon \right) \; .\nonumber\\
\label{eff_lagrangian}
\end{eqnarray}

As before, the expectation value of all quantities are denoted $\epsilon_0$, $\omega_0$, {\it etc.}.
The equation of motion for $\omega$ is
\begin{equation}
  2 \; \frac{\partial}{\partial \omega} \, {\cal S}_{\rm eff} =
  - i \, \epsilon_0 + m^2 + \lambda (\omega_0 + \vec{\sigma}_0^2 ) = 0 \; .
  \label{eom_omega}
\end{equation}
We introduce the effective mass,
\begin{equation}
  m^2_{\rm eff} = i \, \epsilon_0 = m^2 + \lambda (\omega_0 + \vec{\sigma}_0^2) \; .
  \label{meff}
\end{equation}

The $\widetilde{\omega}$ field only appears in two places, and so its equation of motion,
\begin{equation}
  2 \; \frac{\partial}{\partial \widetilde{\omega}} \, {\cal S}_{\rm eff} = - i \widetilde{\epsilon}_0 + Z = 0 \; ,
  \label{tilde_omega_eom}
\end{equation}
just fixes $i \, \widetilde{\epsilon}_0 = Z$.

The equation of motion for $\vec{\sigma}$ is that of Eq. (\ref{eom}),
\begin{equation}
  \frac{\partial}{\partial \vec{\omega}} \, {\cal S}_{\rm eff} =
  \left( \frac{1}{M^2} (- \partial_i^2)^2 + Z (-\partial^2) + m_{\rm eff}^2 \right) \vec{\sigma}_0 = 0 \; ,
  \label{sigma_eom_gen}
\end{equation}
where we use the definition of $m_{\rm eff}$.

It is useful to introduce the propagator for the static, transverse mode.  In momentum space, the static inverse
propagator for $\vec{\chi}$ is, suppressing the isospin indices,
\begin{equation}
  \Delta_{\chi}^{-1}(E = 0, \vec{k}) 
  =\frac{1}{M^2} (\vec{k}^2)^2 + Z \, \vec{k}^2 + m^2_{\rm eff} \; .
  \label{inverse_chi_prop}
\end{equation}
We have used Eq. (\ref{eom_omega}) to fix $\epsilon_0$ in terms of $m^2_{\rm eff}$, and Eq. (\ref{tilde_omega_eom})
to set $\widetilde{\epsilon}_0$.

The equation of motion for $\epsilon$ is
\begin{equation}
  2 \, i \; \frac{\partial}{\partial \epsilon} \, {\cal S}_{\rm eff} =
  - \omega_0 + (N-2) \;  \int \frac{d^3k}{(2 \pi)^3} \; \Delta_\chi (0,\vec{k}) = 0 \; .
  \label{eom_epsilon}
\end{equation}
while that for $\widetilde{\epsilon}$ is
\begin{equation}
  2 \, i \; \frac{\partial}{\partial \widetilde{\epsilon}} \, {\cal S}_{\rm eff} =
  - \widetilde{\omega}_0 + (N-2) \; \int \frac{d^3k}{(2 \pi)^3} \; k^2 \; \Delta_\chi (0,\vec{k}) = 0 \; .
  \label{eom_epsilon_tilde}
\end{equation}
Remember this is only for the static mode at nonzero temperature, and so a factor of temperature has
been absorbed into the couplings and fields.  At zero temperature, the corresponding integral is then over all four momenta,
$E$ and $\vec{k}$.

We can now make a straightforward analysis for the ansatz of a chiral spiral with a single mode,
Eq. (\ref{cs_ansatz}).  The equation of motion for the $\vec{\sigma}$ field, Eq. (\ref{sigma_eom_gen}), gives
\begin{equation}
  \frac{1}{M^2} \, k_0^4 + Z \, k_0^2 + m^2_{\rm eff} = 0 \; ,
\label{cond_k0}
\end{equation}
or $\vec{\sigma}_0 = 0$.  Now it is useful to recognize a remarkable fact, that the inverse propagator
in Eq. (\ref{inverse_chi_prop}) has {\it exactly} the same structure as Eq. (\ref{cond_k0}).  
Consequently, the (static) inverse transverse propagator vanishes at $k_0$, $\Delta^{-1}_\chi(0,k_0 \, \hat{\vec{k}}) = 0$.

This still does not fix the value of $k_0$.  However, for a physical field
the propagator must be positive everywhere or the
theory is unstable.  It is automatically positive at large $k$, while positivity at $k=0$ implies that
$m^2_{\rm eff} > 0$.  As a quadratic equation in $k_0^2$,
there can be two roots of Eq. (\ref{cond_k0}), at $k_0^-$ and $k_0^+$.
If $k_0^- < k_0^+$, the propagator is positive for $k < k_0^-$, 
crosses zero at $k_0^-$, negative when $k_0^- < k < k_0^+$, and positive again when $k> k_0^+$.
If $k_0^- \neq k_0^+$, then, the propagator is negative for some range of momenta, and the theory is unstable.

The only way that Eq. (\ref{cond_k0}) can be satisfied is if there is only one zero for the inverse
propagator, with $k_0^- = k_0^+$.  This implies that the propagator is extremal at $k_0$:
\begin{equation}
  \left. \frac{\partial}{\partial k} \Delta^{-1}(0, k \, \hat{\vec{k}})\right|_{k = k_0} =
  \frac{4}{M^2} \left( k_0^2 + \frac{Z}{2} \, M^2 \right) k_0 = 0 \; .
  \label{cond_vary_k0}
\end{equation}
This is the same equation as we obtained by varying the Lagrangian, evaluated for the chiral spiral, with
respect to $k_0$ in Eq. (\ref{soln_k0}).  Since the propagator is extremal at $k_0$, 
the inverse propagator has a double zero at $k_0$,
\begin{equation}
  \Delta_\chi^{-1}(0,\vec{k}) = \frac{1}{M^2} \left( \vec{k}^2 - k_0^2 \right)^2 \; .
  \label{inverse_chi_prop_solve}
\end{equation}
In other words, just by requiring that the theory has a transverse propagator which is everywhere positive,
that and the equations of motion {\it force} a double pole in $\Delta_\chi(0,\vec{k})$ at $k = k_0$.

This is precisely the same conclusion as we found perturbatively in Secs. (\ref{sec:single}) and (\ref{sec:general}).
While before we could only suggest that the tadpole integral over the transverse propagator gives a linear
infrared divergence, Eq. (\ref{tadpole_div}), at large $N$ these integrals are forced upon us by
Eqs. (\ref{eom_epsilon}) and (\ref{eom_epsilon_tilde}), and there is no escape from a linear infrared divergence.

At large $N$, the $N-2$ fluctuations must be included in order to obtain a self consistent solution, assuming
that the quartic coupling $\lambda N$ is held fixed as $N \rightarrow \infty$.  We demonstrate in the next section
that there is a self consistent solution for the theory, as a phase disordered by quantum fluctuations.

\section{Explicit solution at large $N$}
\label{sec:soln_largeN}

In the previous section we showed that assuming a chiral spiral with a single mode produces a double
pole in the propagator for the transverse mode.  We demonstrate in this section that there is a
nonperturbative solution for the symmetric phase, even when mean field indicates a condensate of
chiral spirals, when $m_{\rm eff}^2$ and $Z$ are negative.
We starts with the Lagrangian of Eqs.~(\ref{lag0}) and (\ref{laghd}),
and for simplicity assume $1/M_1 = 1/M_2 = \kappa = 0$.  It is trivial to generalize our analysis to the
general case.
Since the constraint field for $(\partial_i \vec{\chi})^2$ in Eq. (\ref{constraint_two}) did not make
a significant difference, we ignore it, and only introduce $\omega = \vec{\phi}^2$, Eq. (\ref{constraint_one}),
and integrate over all $\vec{\phi}$.  
If the six-point coupling $\kappa$ vanishes, it is possible to integrate over $\omega$ and obtain
an effective action only in terms of the constraint field, $\epsilon$:
\begin{equation}
\label{eff_lagrangian1}
  {\cal S} = \int d^3 x \; \frac{\epsilon^2}{4 \lambda}
  + \frac{N}{2} \, {\rm tr} \log \left(  \frac{1}{M^2} (-\partial^2 )^2 
    + Z (- \partial^2) + m^2 - i \epsilon \right) \; .
\end{equation}
We expand about a saddle point
\begin{equation}
  \epsilon = i \epsilon_0 + \epsilon_{\rm qu} \; ,
  \label{def_epsilon}
\end{equation}
where $\epsilon_0$ is determined by
\begin{equation}
    \epsilon_0 -
    \frac{\lambda N}{2} \, {\rm tr} \frac{1}{(-\partial^2 )^2/M^2
      + Z (- \partial^2) + m_{\rm eff}^2 } = 0 \; ,
    \label{ep_eom_gen1}
\end{equation}
and we define the renormalized mass,
\begin{equation}
  m_{\rm eff}^2 = m^2 + \epsilon_0 \; .
  \label{define_ren_mass}
\end{equation}
We look for the simplest solution, with constant $\epsilon_0$.  This excludes a chiral spiral
condensate which involves multi-modes, Eq. (\ref{cs_ansatz_multi}).  As we show at the end of this section, however,
our solution is at least locally stable, and so while we cannot exclude it, it appears
unlikely that a multi-mode chiral spiral has a lower action.

For constant $\epsilon_0$, the transverse propagator is
\begin{equation}
  \Delta(\vec{k}) = \frac{1}{(\vec{k}^2)^2/M^2 + Z \vec{k}^2 + m_{\rm eff}^2} \; .
  \label{trans_prop}
\end{equation}
We need to evaluate
\begin{eqnarray}
  \label{tr0}
  {\rm tr} \; \Delta &\equiv&
    \int \frac{d^3 k }{(2 \pi)^3} \; \Delta(\vec{k}) = \frac{M^2}{4 \pi^2} \int\limits_{-\infty}^{+\infty}
    \mathrm d k \, \frac{k^2}{(k^2 + m_{+}^2)(k^2 + m_{-}^2)} 
    \nonumber\\
&=& \frac{M^2}{4 \pi^2} \; \frac{1}{m_+^2 - m_-^2} \int\limits_{-\infty}^{+\infty} \mathrm d k \, 
    \left( \frac{m_+^2}{k^2 + m_+^2} - \frac{m_-^2}{k^2 + m_-^2} \right) \; .\\ \nonumber
\end{eqnarray}
where 
\begin{equation}
  m_{\pm}^2 =  \frac{Z M^2}{2} \left(1 \pm \sqrt{1 - \alpha^2 }\right) \; \; ;\;\;
  \alpha = \frac{2 \, m_{\rm eff} }{ |Z| \, M} \; .
  \label{soln_masses}
\end{equation}  
To evaluate the integral it is necessary to take care with where the poles lie in the complex $k$ plane.

\subsection{Solution for the effective mass}
\label{sec:soln_eff}

We start with the case where $\alpha \leq 1$, so that $m^2_{\pm}$ is real and positive, so we can
assume the same for $m_+$ and $m_-$.  Eq. (\ref{tr0}) equals
\begin{equation}
  {\rm tr} \; \Delta \; = \; \frac{M^2}{4 \pi} \; \frac{1}{m_+ + m_-} \; .
  \label{integral_delta}
\end{equation}
As $m_+^2$ and $m_-^2$ are solutions to a quadratic equation in $k^2$,
$(k^2)^2 + Z M^2 k^2 + m^2_{\rm eff} M^2 = (k^2 + m_+^2)(k^2 + m_-^2) = 0$,
\begin{equation}
  (m_+ + m_-)^2 = 2\, \sqrt{m_+^2 m_-^2} + m_+^2 + m_-^2  = 2 \, m_{\rm eff} M + Z \, M^2 \; .
  \label{mass_identity}
\end{equation}
This is valid for positive $m_{\rm eff}$ and either sign of $Z$, {\it if} the quantity
$2 m_{\rm eff} + Z M$ is positive.
The saddle point equation of Eq. (\ref{ep_eom_gen1}) becomes
\begin{equation}
  m^2_{\rm eff} - m^2 = \lambda_0 \; \frac{M^{3/2}}{\sqrt{2 \, m_{\rm eff} + Z \, M }} \; ,
\label{saddle_explicit1}
\end{equation}
where we define the rescaled quartic coupling
\begin{equation}
  \lambda_0 = \frac{\lambda N }{8\pi}  \; .
\end{equation}
At large $N$, $\lambda_0$ and all other quantities, $m_{\rm eff}$, $m$, $M$, and $Z$,
are of order one as $N\rightarrow \infty$.
The mass dimensions also match, as in three dimensions $\lambda$ has dimensions of mass.

For given values of $Z$, $m^2$ and $\lambda_0$, we need to determine the solution for $m^2_{\rm eff}$.
Notice that  the left hand side of Eq. (\ref{saddle_explicit1})
is a monontonically increasing function of $m_{\rm eff}$, while the right hand side is a monotonically
decreasing function.  The solution is then just the intersection of the two.  The solution can be
found numerically for a given value of the parameters.

It is useful to consider various limits.  At large, positive $Z$, the effective mass is
\begin{equation}
  m^2_{\rm eff} \approx m^2 + \, \frac{\lambda_0 \, M }{\sqrt{Z}} + \ldots \; \; ; \;\; Z \rightarrow + \infty \; .
  \label{soln_large_pos_Z}
\end{equation}
Large $Z$ suppresses fluctuations, with the correction to the bare mass $\sim \lambda/\sqrt{Z}$.  

As $Z$ decreases, the correction to the effective mass grows.  For simplicity, we begin
with the line where
the bare mass vanishes, $m^2 = 0$.  Letting $Z$ decrease, there is a point where $\alpha = 2 m_{\rm eff}/(Z M) = 1$; from
Eq. (\ref{saddle_explicit1}), this happens when
\begin{equation}
  Z_{1}(0) = \left(2^{3/2} \frac{\lambda_0}{M}\right)^{2/5} \; \; ; \; \; m_{\rm eff} =
  \left( \frac{\lambda_0^2 M^3}{4} \right)^{1/5} \; .  
  \label{soln_alpha1_m2_zero}
\end{equation}
For arbitrary $m^2$, we denote the point at which $\alpha = 1$ as $Z_1(m^2)$.

For constant $m^2 = 0$, as $Z$ decreases below $Z_1(0)$ nothing particularly interesting happens
for $m_{\rm eff}$.  For example, when both $Z$ and $m^2$ vanish the effective mass remains nonzero,
\begin{equation}
  m_{\rm eff} = \left( \frac{\lambda_0^2 M^{3}}{2} \right)^{1/5} \;\; ; \;\;
  Z = m^2 = 0 \; .
  \label{soln_zero_Z_m2}
\end{equation}
Clearly this behavior, $m_{\rm eff} \sim \lambda_0^{2/5}$, arises from a nontrivial resummation of
perturbation theory at large $N$.  

As $Z$ decreases to negative values for $m^2 = 0$, mean field theory suggests that the theory exhibits
a condensate of chiral spirals.  We demonstrated in the previous section, however, that such a condensate exhibits
double poles for the transverse modes, Sec. (\ref{sec:gen_negZ}), which disorder the condensate.

However, there is {\it always} a self-consistent solution for the symmetric phase,
Eq. (\ref{saddle_explicit1}); it is only necessary is to ensure that the quantity $2 m_{\rm eff} + Z M$ is positive.
This is easy to do: for example, when $Z$ is large and negative, the solution of Eq. (\ref{ep_eom_gen1}) is
\begin{equation}
  m_{\rm eff} = - Z \, \frac{M}{2} + \frac{16 \lambda_0^2}{M} \, \frac{1}{Z^4} + \ldots \;\; ; \;\;
  Z \rightarrow - \infty \; .
  \label{soln_large_neg_Z}
\end{equation}

Thus along the entire line of $m^2 = 0$, for both positive and negative values of $Z$, the theory is in the symmetric phase.
This remains valid when $m^2$ is positive.  The point where $\alpha = 1$ changes with $m^2$, where
$Z_1(m^2)$ is a monotonically increasing function of $m^2$.

\begin{figure}[hbt!]
  \includegraphics[width=0.9\linewidth]{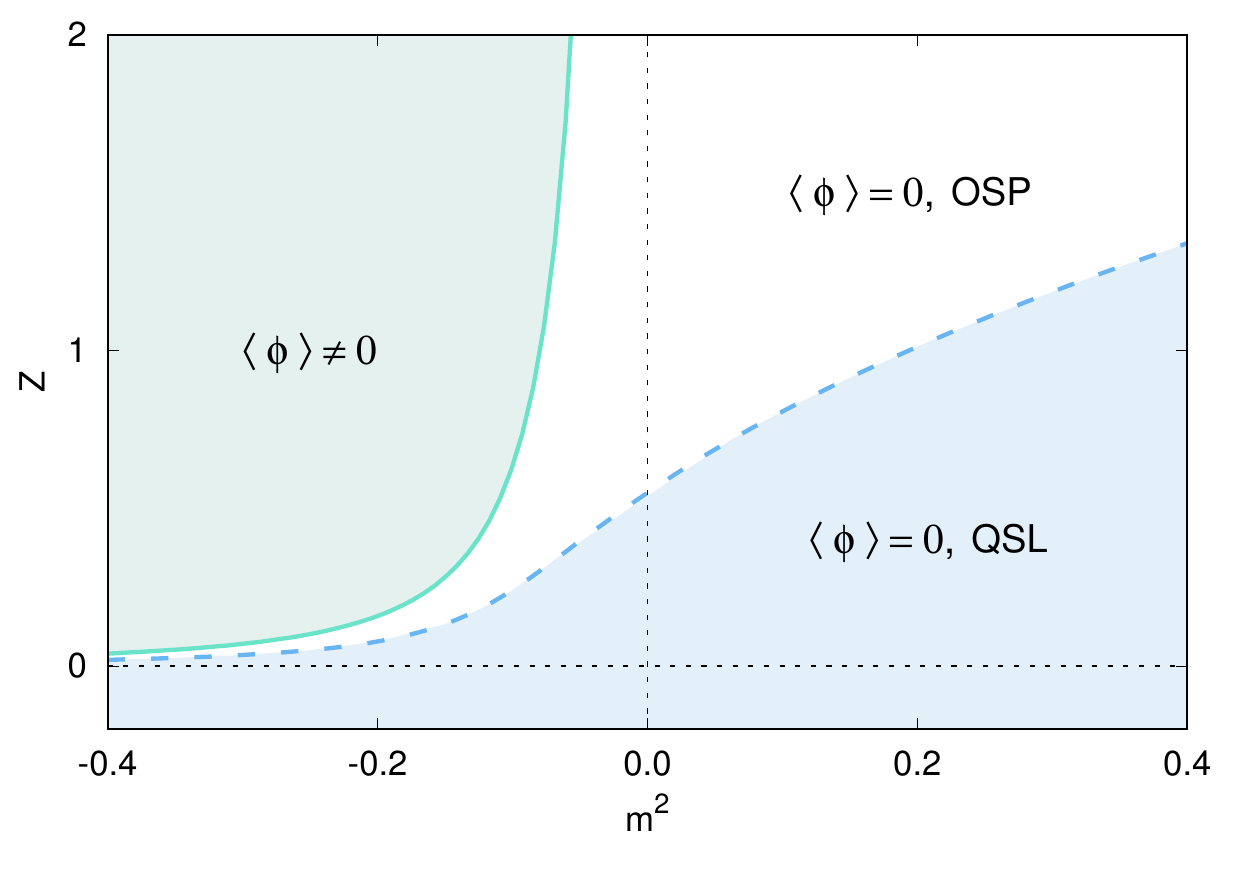}
  \caption{The phase diagram at large $N$, in the plane of the bare mass squared, $m^2$, and the wave function renormalization
    constant, $Z$, for positive quartic coupling $\lambda = 1$.  In terms of the global order parameter, there is
    just the broken phase, where $\langle \phi \rangle \neq 0$, and the symmetric phase, where
    $\langle \phi \rangle = 0$.  Nevertheless, there are two parts to the symmetric phase:
    an ordinary symmetric phase (OSP) and the
    quantum spin liquid (QSL).  In the OSP the two point function of $\phi$ is the usual exponential,
    Eq. (\ref{two_pt_sym}), while in the QSL it also oscillates, Eq. (\ref{two_pt_qsl}).  
    The solid line in the upper left quadrant is a line of second order phase transitions.
    The dashed line in the upper half plane is
    the boundary between the OSP and the QSL, which is not a phase transition.}
  \label{fig:phase_diagram_lambda_positive}
\end{figure}

When $m^2$ becomes negative, the theory enters the broken phase for a fixed, positive value of $Z$.
Since the quartic coupling $\lambda$ is positive, this is a second order transition, determined
by the condition that the effective mass vanishes.  The critical value of the bare mass is given by 
setting $m^2_{\rm eff} = 0$ in Eq. (\ref{saddle_explicit1}), which is just
\begin{equation}
  m^2_{\rm crit} = - \; \frac{\lambda_0 M}{\sqrt{Z}} \; .
  \label{cond_crit_mass}
\end{equation}
For $Z > 0$ and $m^2 < - m^2_{\rm crit}$, the theory is in the broken phase.  This can be treated following the analysis
of Sec. (\ref{sec:gen_posZ}), but other than algebraic complications, 
there are no surprises.  The broken phase has massless transverse modes,
and a nonzero value for a constant condensate.  The phase diagram which results 
is illustrated in Fig. (\ref{fig:phase_diagram_lambda_positive}).  

In mean field theory, a Lifshitz point occurs when $m^2 = Z = 0$; in the full theory,
the corresponding condition is $m_{\rm eff}^2 = Z = 0$.  However, this requires
that $m^2 = m^2_{\rm crit} \rightarrow -\infty$ as $Z \rightarrow 0^+$.  This is
not a Lifshitz point, but a singular limit.  That there is no true Lifshitz point
agrees with general analysis
\cite{Fradkin:1991nr,Erzan:1977,Sak:1978,Grest:1978,Fredrickson:2006,Bonanno:2014yia,Zappala:2017vjf,Zappala:2018khg,Pisarski:2018bct}.
We also discuss in the next section why $m^2_{\rm crit} \rightarrow - \infty$ as $Z \rightarrow 0^+$.

\subsection{Quantum spin liquid}
\label{sec:qsl}

The solution for the effective mass obscures interesting physics associated with how the masses $m_+$
and $m_-$ change as the bare mass and $Z$ are varied.

We assume that $m^2$ is positive and fixed, and vary $Z$.  The extension
to negative $m^2$ trivial; it is just necessary to recognize the transition to the broken phase in
Fig. (\ref{fig:phase_diagram_lambda_positive}).

We start at large, positive $Z$, where $\alpha \leq 1$.  The masses squared in Eq. (\ref{soln_masses}) are real,
so we can take $m_+$ and $m_-$ to be real.  The poles of the propagator are for
$k^2 = - m_\pm^2$, and so along the imaginary axis, at $k = \pm i m_+$ and $k = \pm i m_-$.
The two point function of the scalar field is a sum of exponentials,
\begin{equation}
  \left. \langle  \phi^i(x) \phi^j(0) \rangle\right|_{x \rightarrow \infty}
  = \delta^{ i j} \left( c_- \, {\rm e}^{- m_- x} + c_+ \, {\rm e}^{- m_+ x} \right) \; \; ; \;\;
  \alpha \leq 1 \; .
  \label{two_pt_sym}
\end{equation}
At large $Z$, $m_-$ is light, $m_- \approx m/\sqrt{Z}$, and $m_+$ heavy, $m_+ \approx \sqrt{Z} M$.  Of course
at large distances the light excitation dominates.

As $Z$ decreases from large, positive values, both poles
remain on the imaginary axis, as $m_+$ decreases and $m_-$ increases.  At the point where
$Z = Z_1(m^2)$, these poles merge, with
\begin{equation}
  m_+ = m_- = m_0 \equiv \sqrt{m_{\rm eff} \, M} \; .
  \label{soln_m0}
\end{equation}
When $Z < Z_1(m^2)$, the masses squared of $m_\pm^2$ have both real and imaginary parts, and the poles of
the propagator are at
\begin{equation}
  k = \exp\left( \pm i \left(\frac{\pi}{2} \pm' \frac{\theta}{2}\right) \right) \; m_0 \; \; ; \;\; Z \geq 0 \; ,
  \label{positions_poles}
\end{equation}
where
\begin{equation}
  \tan (\theta) = \sqrt{\alpha^2 - 1} \;\; ; \;\; \alpha \geq 1 \; .
  \label{def_theta}
\end{equation}
There are four poles, one in each quadrant.

As $Z$ decreases below $Z_1(m^2)$, the poles develop both real and imaginary parts.  Defining 
\begin{equation}
  m_r = m_0 \, \cos\left(\frac{\theta}{2}\right) \;\; ; \;\; m_i = m_0 \sin\left(\frac{\theta}{2}\right) \; \; ; \;\;
  Z \geq 0 \; ,
\end{equation}
and the two point function is
\begin{equation}
  \left. \langle  \phi^i(x) \phi^j(0) \rangle\right|_{x \rightarrow \infty}
  = \delta^{ i j} \; {\rm e}^{- m_r x} \; \left( c_1 \cos(m_i x ) + c_2 \, \sin(m_i x) \right)
\; \; ; \;\; \alpha \geq 1 \; .
  \label{two_pt_qsl}
\end{equation}
Because the poles have a nonzero real part, the two point function is the usual exponential of $m_r x$
times cosine or sine of $m_i x$.  For positive $Z$ less than $Z_1(m^2)$, $m_r > m_i$.  

As $Z \rightarrow 0^+$, by Eq. (\ref{soln_zero_Z_m2}) the effective mass is nonzero,
so $\alpha \rightarrow \infty$ and $\theta = \pi/2$.  The four poles are then at $\pm \pi/4$
and $\pm 3 \pi/4$, with $m_r = m_i = m_0/\sqrt{2}$; $m_0$ is given by Eqs. (\ref{soln_zero_Z_m2}) and (\ref{soln_m0}).

For negative $Z$ the poles are at
\begin{equation}
  k = \exp\left( \pm  \frac{i\theta}{2} \right) \; m_0 \; \; ; \;\;
  k = \exp\left( \pm i \left(  \pi -  \frac{\theta}{2} \right) \right) \; m_0 \; \; ; \;\; Z \leq 0 \; .
  \label{positions_poles_negZ}
\end{equation}
Due to the overall factor of sign of $Z$ in Eq. (\ref{soln_masses}),
when $Z$ changes sign the poles rotate by $\pi/2$.  The limit of $Z \rightarrow 0^\pm$ is consistent as
then the poles are spaced by $\pi/2$ anyway.

For negative $Z$ the two point function remains an exponential times oscillatory functions, but now the
expressions for $m_r$ and $m_i$ become
\begin{equation}
  m_r = m_0 \, \sin\left(\frac{\theta}{2}\right) \;\; ; \;\; m_i = m_0 \cos\left(\frac{\theta}{2}\right) \; \; ; \;\;
  Z \leq  0 \; ,
\end{equation}

As $Z$ decreases from zero, $\alpha$ decreases, although it must remain greater than unity.  
As $Z \rightarrow - \infty$, for example, from Eqs. (\ref{soln_large_neg_Z}),
(\ref{soln_m0}), and (\ref{def_theta}), $m_0 \sim |Z|^{1/2}M/\sqrt{2}$ and
$\theta \sim 8 \lambda_0/(M |Z|^{5/2})$, so that
\begin{equation}
  m_r \approx \frac{ 2^{3/2} \lambda_0}{Z^2} + \ldots \;\; ; \;\;
  m_i \approx  \frac{|Z|^{1/2} M}{\sqrt{2} } + \ldots \; \; ; \;\; Z \rightarrow - \infty \; .
\end{equation}

In mean field theory $m_{\rm eff}$ vanishes, so the poles of the propagator
are at $k^2 = 0$ and $k^2 = Z M^2$.  If $Z < 0$, the latter is on the axis of real, positive $k^2$, with
$m_i = |Z|^{1/2} M$.  This indicates the instability of the ground state to the formation of a condensate
of chiral spirals.  At large $N$, though, instead the theory remains in the symmetric phase.
When $Z$ is large and negative, the poles in $k^2$ are {\it close} to the real axis, with
$m_i \sim |Z|^{1/2} M/\sqrt{2}$, but the phase is stable,
as each pole also has a small, real part, $m_r \sim \lambda_0/Z^2$.
As $Z$ becomes more negative,
the two point function oscillates over distances which are shorter and shorter relative to
the distance over which it falls exponentially, $1/m_i \ll 1/m_r$.  This is the signal for an
unusual form of disorder, which we term a quantum spin liquid
\cite{Fradkin:1991nr,Ioffe:1988a,Zhou:2016cnl,Savary:2016ksw,Kharkov:2017cvm,Kharkov:2018lri,Kharkov:2019fvv,OBrien:2020bho}.

The boundary between the usual symmetric phase, with the two point function as in Eq. (\ref{two_pt_sym}),
and a quantum spin liquid, with that of Eq. (\ref{two_pt_qsl}), is indicated by a dotted line
in Fig. (\ref{fig:phase_diagram_lambda_positive}).  This line is the curve
$Z_1(m^2)$, where $\alpha =1$.  The includes the entire region where $Z < 0$, plus a region
where $Z$ is small and positive, $0< Z < Z_1(m^2)$.  This is in contrast to mean field theory,
where there is a condensate of chiral spirals only for $Z< 0$ and $m^2 < m_1^2$, where
$m_1^2 > 0$; see Fig. (1) of Ref. \cite{Pisarski:2018bct}.  This difference is due to quantum fluctuations at large $N$.

While the behavior of the two point function changes as one goes from the ordinary symmetric phase
into a quantum spin liquid, in our model this is not a phase transition.  We have concentrated on
the static mode at nonzero temperature.  For a static mode with mass squared $m^2$, it contributes
to the free energy as
\begin{equation}
  {\cal F}(m) \sim T \int d^3 k \; \left(
    {\rm tr}  \log \left( k^2 + m^2 \right) - \frac{m^2}{k^2} \right) \sim T  \left( m^2 \right)^{3/2} \; .
\end{equation}
In the present case, we have several masses which contribute.  For $\alpha \leq 1$ both are real.
When $\alpha \geq 1$ there are four complex masses, but 
they are always paired into complex conjugates, so that the sum is real.  Consequently, the free energy,
and any finite number of derivatives thereof, behave smoothly as $\alpha \rightarrow 1$.

Logically, it is possible that the imaginary part of the mass, $m_i$, vanishes at the same time as the real part, $m_r$;
if so, this might produce a novel
critical point.  From the phase diagam of Fig. (\ref{fig:phase_diagram_lambda_positive}), though, this does not
occur: the dotted line, separating the ordinary symmetric phase and the quantum spin liquid, never
intersects the line of second order phase transitions.  

This also helps to understand one feature of the phase diagram, which is that 
the bare mass squared diverges, $m^2_{\rm crit} \rightarrow - 1/\sqrt{Z}$, as $Z \rightarrow 0$.  This occurs
because the broken phase cannot exist for negative $Z$, so the transition to the symmetric phase
arises for a small but nonzero value of $Z$.

\subsection{Stability of the quantum spin liquid phase}
\label{sec:qsl_stability}

While we have established that there is a novel solution for the symmetric phase when $\alpha > 1$, we also
need to show that it is stable.  This is actually direct.  From the expansion of $\epsilon$ in Eq.
(\ref{def_epsilon}), the two point function of $\epsilon_{\rm qu}$ is, in momentum space,
\begin{equation}
  \langle  \epsilon_{\rm qu}(\vec{p}) \epsilon_{\rm qu}(-\vec{p}) \rangle =
  + \frac{1}{2 \lambda} + N 
  \int \frac{\mathrm d^3 k}{(2\pi)^3} \, \Delta(\vec{k}) \Delta(\vec{k} + \vec{p} ) \; ,
  \label{stability_eps}
\end{equation}
where $\Delta$ is the transverse propagator of Eq. (\ref{trans_prop}).  
Notice that the sign of the second term is positive in Eq. (\ref{stability_eps}) because the contour
of integration for $\epsilon_{\rm qu}$ runs along the real axis.  This is standard in a large $N$ expansion
\cite{Tsvelik:1996zj,Moshe:2003xn}.

It is also direct to compute the propagator for the $\widetilde{\epsilon}$ field of Eq. (\ref{eff_lagrangian}):
\begin{equation}
  \langle  \widetilde{\epsilon}_{\rm qu}(\vec{p}) \widetilde{\epsilon}_{\rm qu}(-\vec{p}) \rangle =
  + \frac{1}{2 \lambda} + 4 N 
  \int \frac{\mathrm d^3 k}{(2\pi)^3} \, (\vec{k} \cdot ( \vec{k} + \vec{p}))^2
  \; \Delta(\vec{k}) \; \Delta(\vec{k} + \vec{p} ) \; ,
  \label{stability_eps_tilde}
\end{equation}
Because of the complicated pole structure of the propagators, it is involved to explicitly evaluate these
inverse propagators.  However, both Eqs. (\ref{stability_eps}) and (\ref{stability_eps_tilde}) are convergent
in both the ultraviolet and infrared limits.  
Since the propagator $\Delta \sim 1/(\vec{k}^2)^2$ at large momentum, ultraviolet convergence
follows directly by power counting.  Infrared convergence
is guaranteed because the inverse propagator for our solution is always gapped.  Thus both propagators
are convergent integrals over positive quantities, and so are also positive.  This implies local stability.

This is important because when $Z < 0$. it is possible that at large $N$ there is a 
multi-mode solution for a chiral spiral, Eq. (\ref{cs_ansatz_multi}).  We expect that such a solution
has a double pole at $k_0$, but cannot prove this.  However, the quantum spin liquid solution is at least
locally stable, and it seems very likely that it is the global minimum.

\section{Phase diagram including transverse fluctuations}
\label{sec:phase_diagram}

In mean field theory, there are three phases: symmetric, broken,
and one with a spatially anisotropic condensate \cite{Pisarski:2018bct}.
For now we allow $N$ to be arbitrary, so the latter can be either a kink crystal, for $N=1$, or a chiral
spiral, for $N \geq 2$.  The transition between the symmetric and broken
phases is of second order (when the quartic coupling $\lambda > 0$); that between the symmetric and chiral spiral phases is of
second order; and that between the the broken and chiral spiral phase is first order, Fig. (1) of Ref. \cite{Pisarski:2018bct}.

Fluctuations \cite{Brazovski:1975,Ling:1981,Dyugaev:1982gf,Hohenberg:1995,Karasawa:2016bok} turn the transition between
the symmetric and chiral spiral phases into a line of first order transitions, Fig. (2) of Ref. \cite{Pisarski:2018bct}.
This is due to fluctuations in the longitudinal mode, whose propagator is
\begin{equation}
  \Delta_{\rm long}(\vec{k}) \sim \frac{1}{\delta k_z^2 + (4 \, k_0 \, \delta k_z \, 
    + \widetilde{k}^{\; 2})\widetilde{k}^2/M^2 + m^2} \; .
  \label{long_prop}
\end{equation}
Here we assume that the condensate is along the $z$ direction, $k_0$
the characteristic momentum of the condensate, $\delta k_z = k_z - k_0$
and $\widetilde{k}^{\; 2} = (\delta k_z)^2 + k_\perp^2$.
Notably, there are no terms quadratic in $k_\perp$.  In mean field theory,
the transition between the symmetric and spatially anisotropic condensate phase occurs when $m^2 = 0$.
As pointed out by Brazovski \cite{Pisarski:2018bct,Brazovski:1975,Ling:1981,Dyugaev:1982gf,Hohenberg:1995,Karasawa:2016bok},
this gives rise to a linear infrared divergence.  The transition occurs as the parameters $m^2$ and $Z$ are
changed, so $m^2$ jumps from one nonzero value to another, through a first order transition.

This is very similar to the mechanism proposed in this paper,
by which (static) transverse fluctuations disorder a chiral spiral when $N \geq 2$.  
We find that the would be Goldstone bosons have a double pole when $k = k_0$, although not along the direction
of the condensate.  This produces a tadpole diagram which is linearly divergent in the infrared, Eq. (\ref{tadpole_div}).
This occurs throughout the chiral spiral phase, however.  As we demonstrated in the previous section,
Sec. (\ref{sec:soln_largeN}), at large $N$ the transverse fluctuations disorder the chiral spiral phase, with a
propagator which is
\begin{equation}
  \Delta^{i j}(E,\vec{k}) = \frac{\delta^{i j} }{ E^2 + (\vec{k}^2)^2/M^2 + Z \vec{k}^2 + m_{\rm eff}^2} \; .
  \label{sym_prop}
\end{equation}
The isospin indices are $i$ and $j$, so the propagator is symmetric.  At large $N$ we showed that even in a region
where one expects a condensate of chiral spirals --- when $m^2$ and $Z$ are both negative --- that at nonzero
temperature static transverse fluctuations disorder the system.  

At large $N$ we find that there is no phase transition between the ordinary symmetric phase and
that with a quantum spin liquid, only between the ordinary symmetric phase and the broken phase,
Fig. (\ref{fig:phase_diagram_lambda_positive}).

While our conclusions are only certain at large $N$, we suggest that they hold for {\it any} $N \geq 2$.
Consider the effective Lagrangian of Eq. (\ref{eff_lagrangian}), and consider an expansion about $N = 2 + \delta$,
where $\delta \ll 1$.  In this instance, we can expand in $\delta$.  The leading terms are given by the classical
theory, and quantum fluctuations about that.  However, the transverse terms, $\sim \delta$, inevitably bring in
the infrared divergences of the would be Goldstone modes, and will disorder the system.

This does not exclude the possibility that the theory is disordered for $\delta \ll 1$, ordered for some intermediate
range of $N$, from $N = 3$ to $N_{\rm max}$, and then disordered again for $N > N_{\rm max}$.  This can be studied
most directly through numerical simulations of the scalar theory.  One of us has performed numerical simulations
for $N = 8$ and $10$ which indicate there is no phase with a condensate of chiral spirals \cite{valgushev}.
In Sec. (\ref{sec:NJL}), we suggest a Nambu Jona-Lasino model, where $N=3$, where this could also be analyzed
at nonzero chemical potential.

These numerical simulations \cite{valgushev} show that even if there is a $N_{\max}$, it is difficult to distinguish
between a standard symmetric phase and a condensates of chiral spirals.  If there is a line of first
order transitions between the two phases, as predicted by the analysis of
Brazovski \cite{Pisarski:2018bct,Brazovski:1975,Ling:1981,Dyugaev:1982gf,Hohenberg:1995,Karasawa:2016bok}, that
simplifies things greatly.

In all of our analysis we have assumed that the $O(N)$ symmetry is exact.  If
this symmetry is broken by a small but nonzero background field, $h\neq 0$ in Eq. (\ref{lag0}),
even for small $h$, it is not trivial solving for the explicit form of the chiral spiral.  The assumption
which simplified the analysis so greatly, that $\vec{\phi}^2$ is constant, no longer holds.  It is then
like the case of a kink crystal, where it is necessary to solve a nonlinear differential equation to
determine the form of how a deformed chiral spiral, for small $h$, goes over to a kink crystal, for large $h$.

Even so, for small $h$ it is very natural to assume that the form of the propagator is like that of Eq. (\ref{sym_prop}):
for either sign of $Z$, the propagator is always gapped, with
$m_{\rm eff}^2 \neq 0$.  Thus 
for small $h$, it is reasonable to expect that the quantum spin liquid exists for a large range of parameter
space.

We have concentrated on the static mode at nonzero temperature.  At zero temperature, the propagator for
the transverse mode about a chiral spiral, for the tadpole diagram analogous to Eq. (\ref{tadpole_div}),
\begin{equation}
  \sim  \int_{\-\infty}^{+\infty} \frac{dE}{2 \pi} \int \frac{d^3k}{(2 \pi)^3}
  \frac{1}{E^2 + (\vec{k}^2 - k_0^2)^2/M^2} \sim \int \frac{d^3k}{(2 \pi)^3}
  \frac{M}{|\vec{k}^2 - k_0^2|} \; ,
  \label{zero_temperature}
\end{equation}  
after integrating over the energy, $E$, which is continuous at zero temperature.  This has a
logarithmic infrared divergence about $k = k_0$, which mildly washes out a chiral spiral condensate.
This disorder is precisely analogous to that which is expected for the phonon of the longitudinal mode
\cite{chaikin:2010,Fradkin:1991nr,Pisarski:2018bct,Kleinert:1981zv,Kolehmainen:1982jn,Baym:1982ca,Bunatian:1983dq,Takatsuka:1987tn,Migdal:1990vm}.
This suggests that there is a quantitative difference between the quasi-long-range order at zero temperature,
and the quantum spin liquid at nonzero temperature.
We suspect that there is a true phase transition at a small but nonzero temperature, but we have not analyzed this in detail.

\section{Implications}
\label{sec:conclusions}

We have established that static transverse modes disorder a condensate of chiral spirals with a single mode.
Our analysis can be tested by numerical simulations of the scalar field theory.  It is of interest to
study these effects for fermions at nonzero density.  We discuss two examples.

\subsection{Four fermion models}
\label{sec:NJL}

In $1+1$ dimensions, Gross-Neveu models are given by
\begin{equation}
  {\cal L} = \overline{\psi} \!\!\not \! \partial \, \psi
  + g_1 \left( \overline{\psi} \psi \right)^2
    + g_2 \left( \overline{\psi} \gamma_5 \psi \right)^2  \; .
      \label{gross_neveu}
\end{equation}
We take $N_f$ flavors of two component fermions, with an implied sum over flavors:
$\overline{\psi} \psi \equiv \sum_{i = 1}^{N_f} \overline{\psi^i}  \psi^i$, {\it etc.}
The theory is asymptotically free and soluble at large \cite{Gross:1974jv} and 
indeed any \cite{Azaria:2016mqb,James:2017cpc} $N_f$.
When $g_2 = 0$, the generation of mass, $\langle \overline{\psi}\psi\rangle \neq 0$,
spontaneously breaks a $Z(2)$ symmetry.  For the chiral Gross-Neveu model, $g_1 = g_2$,
mass generation spontaneously breaks the global $U(1)$ symmetry of
$\psi \rightarrow \exp(i \theta \gamma_5)$.   The global symmetry is larger than $U(N_f)$, equal to
$O(2 N_f)$ \cite{Dashen:1975xh}, but this symmetry is respected by the dynamical generation of mass.

These theories are soluble at nonzero chemical potential.  
At large $N_f$ the Gross-Neveu model develops a kink crystal \cite{Shei:1976mn,Thies:2006ti,Thies:2019ejd,Thies:2020gfy},
and the chiral Gross-Neveu model, a (multi-mode) chiral spiral \cite{Basar:2008im,Basar:2008ki,Basar:2009fg}.
Numerical simulations of Gross-Neveu models have also been carried out at small $N_f$, and 
support the phase diagram found at large $N_f$
\cite{Pannullo:2019bfn,Pannullo:2019prx,Lenz:2020bxk,Narayanan:2020uqt}.

We note that a type of Gross-Neveu model in $1+1$ dimensions, with two flavors and three colors, is soluble in the limit
of zero bare quark mass, and at small mass by using a truncated conformal spectrum approach
\cite{Azaria:2016mqb,James:2017cpc}.  In the chiral limit, only a discrete $Z(2)$ symmetry is
spontaneously broken.  There is a wealth of phases with quasi-long-range order at nonzero chemical potential,
including the Bose condensation of either scalar mesons or deuterons, and a phase with gapless baryons.

Four fermion theories are non-renormalizable in $2+1$ dimensions, but they can still be considered as a type
of effective theory.  Simulations with the Gross-Neveu model in $2+1$ dimensions \cite{Narayanan:2020uqt}
find a kink crystal, but one which appears to vanish as the lattice spacing goes to zero.

We cannot use a Gross-Neveu model to study our effect, since the global symmetry is not broken.  Indeed, since
Goldstone bosons have logarithmic infrared divergences in two spacetime dimensions, it is probably more useful
to study models in $2+1$ dimensions.  

We suggest the following theory.  Let $\psi$ and $\chi$ represent two component fermions in $2+1$ dimensions,
with again an implicit sum over $N_f$ flavors, $\overline{\psi} \psi = \sum_{i=1}^{N_f} \overline{\psi^i} \psi^i$,
and similarly for $\chi$.  Combine them together as $\zeta^i = (\psi^i, \chi^i)$, with the Lagrangian
\begin{equation}
  {\cal L} = \overline{\zeta} \, \!\! \not \! \partial \, \zeta + g_1 (\overline{\zeta} \zeta)^2 
  + g_0 (\overline \zeta \sigma^a \zeta)^2 \; ,
\end{equation}
where the Pauli matrix $\sigma^a$ acts in the space of $(\psi,\chi)$.
In three spacetime dimensions the mass for a two component fermion is odd under parity and time reversal.
A parity even mass is formed by pairing up 
$\psi$ and $\chi$ together, with masses of equal magnitude and opposite sign \cite{Pisarski:1984dj}.

If $g_0 = 0$, the coupling $g_1$ may spontaneously break the $Z(2)$ symmetry, but not the $U(2 N_f)$
symmetry, and so a dynamically generated mass is odd under parity.
In contrast, if $g_0 = 0$ and $g_1 \neq 0$, it is very possible that the theory spontaneously generates
a parity even mass.  In particular, the term $\overline{\zeta} \sigma^3 \zeta$ tends to give masses of
opposite sign to $\psi$ and $\chi$, and breaks $U(2 N_f)$ to $U(N_f) \times U(N_f)$
\cite{Pisarski:1984dj}. 
For a single flavor $SU(2)$ breaks to $U(1)$, times an overall, unbroken $U(1)$.
One of the $SU(2)$ directions generates a condensate, leaving two Goldstone bosons.
At nonzero density one of the Goldstone bosons can pair with
the condensate to form a chiral spiral, leaving one Goldstone boson to disorder the condensate of
chiral spirals.

This model is presumably soluble at large $N_f$.  We note, however, that the transverse fluctuations
are of $\sim N_f^0$, and so if a chiral spiral condensate arises at infinite $N_f$,
it is only disordered at next to leading order in $1/N_f$.  It may be more useful to use numerical simulations on the lattice,
especially for $N_f = 1$.  This assumes that the physics does not disappear as the cutoff vanishes \cite{Narayanan:2020uqt}.
However, it is easy to add additional, dynamical scalar fields and construct a model
which is both renormalizable and with the same pattern of symmetry breaking, Sec. (4.4) of \cite{Moshe:2003xn}.

As discussed in Sec. (\ref{sec:single}), though, we
expect that the spontaneous breaking of an $O(4)$ symmetry prefers a chiral spiral over a kink crystal.

\subsection{Phase diagram of QCD}
\label{sec:QCD}

Our analysis is relevant for the phase diagram of QCD, in the plane of temperature,
$T$, and the quark chemical potential, $\mu$.  At nonzero chemical potential, it is natural that the effect
of fermion loops turn both the quartic coupling, $\lambda$, and $Z$ negative, Eq. (\ref{lag0})
\cite{Pisarski:2018bct,Overhauser:1960,Migdal:1971,Migdal:1973zm,Migdal:1978az,Kleinert:1981zv,Kolehmainen:1982jn,Baym:1982ca,Bunatian:1983dq,Takatsuka:1987tn,Migdal:1990vm,Kaplan:1986yq,Brown:1993yv,Kojo:2009ha,Kojo:2010fe,Kojo:2011cn,Kojo:2014fxa,Nickel:2009ke,Nickel:2009wj,Buballa:2014tba,Carignano:2014jla,Hidaka:2015xza,Lee:2015bva,Buballa:2015awa,Braun:2015fva,Carignano:2015kda,Heinz:2015lua,Carignano:2016jnw,Azaria:2016mqb,Adhikari:2016jzc,Adhikari:2016vuu,Andersen:2017lre,Adhikari:2017ydi,James:2017cpc,Carignano:2016lxe,Khunjua:2017khh,Khunjua:2017mkc,Andersen:2018osr,Carignano:2018hvn,Buballa:2018hux,Khunjua:2018sro,Khunjua:2018jmn,Carignano:2019ivp,Khunjua:2019lbv,Khunjua:2019ini}.
In particular, when $\lambda$ changes sign, a critical endpoint can arise
\cite{Asakawa:1989bq,Stephanov:1998dy,Stephanov:1999zu,Son:2004iv,Stephanov:2008qz,Parotto:2018pwx,Schaefer:2006ds,Rennecke:2016tkm,Fu:2019hdw,Bzdak:2019pkr}.

It is not clear what the relationship is between the critical endpoint and the region where a chiral spiral arises.
In the simplest Nambu Jona-Lasino models, the critical endpoint coincides with the Lifshitz point
(where $Z = m^2 = 0$) \cite{Buballa:2014tba}, but in general the two are separate \cite{Carignano:2014jla}.
In any case, fluctuations wash out the Lifshitz point
\cite{Fradkin:1991nr,Erzan:1977,Sak:1978,Grest:1978,Fredrickson:2006,Bonanno:2014yia,Zappala:2017vjf,Zappala:2018khg,Pisarski:2018bct}.
Calculations using the Functional Renormalization Group appear to show that the region where $Z < 0$ is large,
while the critical region for the endpoint is small \cite{Fu:2019hdw}.  

In QCD, the order parameter for a chiral spiral only involves the global $U(1)$ symmetry
\cite{Pisarski:2018bct}.  This is not directly affected by our analysis, as a chiral spiral
for $U(1)$ has no transverse modes.  Nevertheless, there is surely a close relation between
the full chiral symmetry and this $U(1)$ subgroup.  The relationship between the two is
involved, and beyond the scope of our analysis here.  

Even so, it is reasonable to conjecture that any region where $Z<0$, and $U(1)$ chiral spirals
arise, affect the propagation of pions and kaons.  We do not expect that the inverse propagator of
pions or kaons vanishes at any point, as that would produce double poles.  It is natural to
conjecture that the dispersion relation of pions and kaons is like that of the symmetric modes in Eq. (\ref{sym_prop}).

This modest assumption has immediate implications.
If the Minkowski energy $\widetilde{E} = i E$, the poles of the propagator in Eq. (\ref{sym_prop}) are
\begin{equation}
  \widetilde{E}(\vec{k})^2 = \frac{(\vec{k}^2)^2}{M^2} + Z \, \vec{k}^2 + m_{\rm eff}^2 \; .
  \label{dispersion_relation}
\end{equation}
Using the standard Bose-Einstein statistical distribution function, such a modified dispersion relation
produces what appears to be deviations from a thermal distribution.  In this case, the system is thermal,
but the dispersion relation is modified.  The effects of a modified
dispersion relation could be quite striking \cite{rdp}.  This is
diluted by integrating over the temperature history and
large boost velocity of the medium.  It is surely necessary not to
look at total abundances, integrated over all momentum,
but as a function of momentum.  A more sensitive probe is to measure
the fluctuations for particles binned with respect to their momentum.  
  
\acknowledgments
R.D.P., A.M.T., and S.V. thank the U.S. Department of Energy for support
under contract DE-SC0012704.  We thank Y. Hidaka for collaborating at the beginning of
this project.  We thank him, B. Friman, and E. Nakano for discussions, and especially
B. Friman for pointing out the work of Kleinert \cite{Kleinert:1981zv} to us.

\bibliography{lifshitz}

\end{document}